\documentclass{PoS}

\usepackage{braket}
\usepackage{subfigure}
\renewcommand{\d}{\ensuremath{\mathrm{d}}}

\title{Many faces of the Landau gauge gluon propagator at zero and finite temperature: positivity violation, spectral density and mass scales}

\ShortTitle{Many faces of the Landau gauge gluon propagator}

\author{\speaker{Paulo J. Silva}, Orlando Oliveira\\
        Center for Computational Physics, University of Coimbra\\
        E-mail: \email{psilva@teor.fis.uc.pt, orlando@teor.fis.uc.pt}}

\author{David Dudal \\
        Ghent University, Department of Physics and Astronomy, Krijgslaan 281-S9, 9000 Gent, Belgium\\
        E-mail: \email{david.dudal@ugent.be}}

\author{Pedro Bicudo, Nuno Cardoso\thanks{Address after September 2013:
NCSA, University of Illinois, Urbana IL 61801, USA.}\\
         CFTP, Instituto Superior T\'{e}cnico, CFTP, Universidade de Lisboa, 1049-001, Lisboa, Portugal\\
        E-mail: \email{bicudo@tecnico.ulisboa.pt, nunocardoso@cftp.ist.utl.pt}
        }

\abstract{We address several aspects of gluon propagation at zero
and finite temperature. In particular, we study the violation of
spectral positivity, we discuss a method to extract the
K\"all\'{e}n-Lehmann spectral density of a particle (be it
elementary or bound state) propagator and apply it to compute
gluon spectral densities from lattice data. Furthermore, we also consider
the interpretation of the Landau gauge gluon propagator at finite
temperature as a massive type bosonic propagator. }

\FullConference{QCD-TNT-III-From quarks and gluons to hadronic matter: A bridge 
too far?,\\
		2-6 September, 2013\\
		European Centre for Theoretical Studies in Nuclear Physics and R
elated Areas (ECT*), Villazzano, Trento (Italy)}

\begin{document}

\section{Gluons at zero temperature}

In recent years, the Landau gauge gluon propagator
\begin{equation}
D^{ab}_{\mu\nu} ( \hat{p} ) ~ = ~ \delta^{ab} ~
   \Big( \delta_{\mu\nu} ~ - ~ \frac{p_\mu p_\nu}{p^2} \Big) ~
   D( p^2 ) ~ \label{propcont}
\end{equation}
has been computed on the lattice, using volumes as large as
$(27~\textrm{fm})^4$ for the SU(2) gauge group \cite{cucc07}
and $(17~\textrm{fm})^4$ for the SU(3) gauge group \cite{bma09}.
This was due to a renewed interest in the infrared behaviour
of the Landau gauge Yang-Mills propagators, in connection with the
gluon confinement phenomenon. Simulations show that the propagators reach
a finite non-zero value in the infrared region. However,
the lattice spacing used in the referred simulations was quite big,
being $0.22$ fm for SU(2) and $0.18$ fm for SU(3).
Despite the large physical volume, the use of such large lattice spacings
changes quantitatively the propagator in the infrared region \cite{olisi12}. 
Although we will not discuss this effect here, it is an important bias, 
together with the Gribov copies effect
\cite{SilOli04,Stern13}, that should not be forgotten. We call
the reader's attention that, in what concerns the ghost propagator, 
the combined effect of lattice spacing and physical volume 
was not investigated so far for the SU(3) gauge group \footnote{For a SU(2) analysis see \cite{lit}.}.

\subsection{Positivity violation of the gluon propagator as a sign of gluon confinement}

It is a well accepted fact that the $\mathcal{S}$-matrix of a non-Abelian gauge theory does not display poles that would correspond to asymptotically observable degrees of freedom with the quantum numbers of gluons (color charged vector particles). This is a simple empirical fact in the case of QCD: we observe no free quarks or gluons, but we do observe pions, mesons etc.~

The strong coupling makes it difficult to address with continuum tools the issue of the nonperturbatively realized QCD spectrum. Useful input can come from gauge fixed lattice simulations of e.g.~the quark and gluon propagator. In this proceeding, we will solely focus on pure glue dynamics and ensuing (Euclidean) gluon propagation. From state-of-the-art lattice simulations \cite{cucc05, aubin04, sioli06, bowman07} in the Landau gauge, a numerical estimate can be obtained for the so-called Schwinger function:
\begin{equation}
  C(t) = \int_{-\infty}^{\infty} \frac{\d p}{2\pi} D(p^2) \exp(-ipt).
\label{schwinger}
\end{equation}
With some complex analysis tools, one can then link $C(t)$ to the K\"all\'en-Lehmann spectral function $\rho(\omega^2)$ of the gluon:
\begin{equation}
  C(t)=  \int_{0}^{\infty} \d\omega \rho(\omega^2) e^{-\omega t},
\label{kallen}
\end{equation}
under the \emph{assumption} the gluon has a standard K\"all\'en-Lehmann representation of the form
\begin{equation}
  D(p^2) = \int_{0}^{\infty} \d \mu \frac{\rho(\mu)}{\mu+p^2}.
\label{specdens}
\end{equation}
As $\rho(\mu)$ has the meaning of a scattering probability, it ought to be positive in a physical Hilbert space. From the correspondence (\ref{kallen}), it is then clear that $C(t)$ should be, at least, also positive.

The gluon Schwinger function $C(t)$ is depicted in Figure 1, clearly displaying a violation of positivity, thence the gluon cannot be attributed a physical meaning. This can be seen as evidence in favour of gluon confinement, see also \cite{cornwall} for more detailed spectral musings.
\begin{figure}[t]
\begin{center}
\vspace*{0.5cm}
\includegraphics[width=0.5\textwidth]{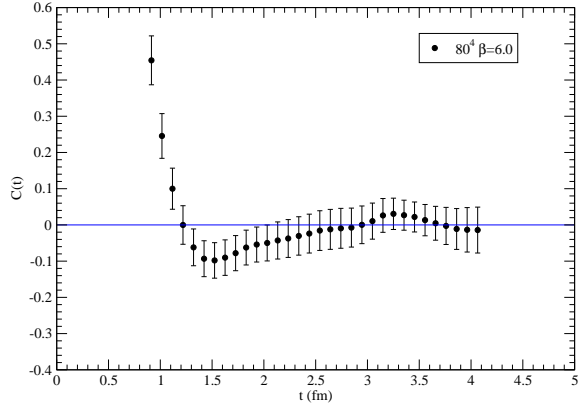}
\caption{Temporal correlator for the gluon propagator computed using $80^4$ $\beta=6.0$ lattice data.}
\end{center}
\end{figure}

\subsection{Determination of the gluon spectral density from lattice data}
The spectral density contains, amongst other things,
information on the masses of physical states described by the
operator $\mathcal{O}$.

We now wish, given data input (with errors) for the propagator at a set of discrete momenta, to obtain a stable estimate for the spectral function. In general, this is a inversion problem. It is interesting to notice that eq.~(\ref{specdens}) is equivalent to applying the Laplace transform twice,
$  D=\mathcal{L}^2\hat\rho=\mathcal{L}\mathcal{L}^\ast \hat\rho$
where $(\mathcal{L}f)(t)\equiv\int_0^{\infty}\d s e^{-st}f(s)$.
This is a notorious ill-posed problem. We used $\mathcal{L}=\mathcal{L}^\ast$.

For positive spectral functions, the inversion can be achieved using the
maximum entropy method (MEM) \cite{mem}. Though, as the gluon Schwinger function already reveals the spectral density cannot be positive over its whole domain, the standard MEM procedure does not apply. We will rely on an alternative approach, preliminary discussed in \cite{latt2012, qchsx, latt2013} with a more complete treatment in \cite{letter}. We found  inspiration in the Tikhonov approach to ill-posed problems, supplemented with the Morozov discrepancy principle. Specifically,
setting $D_i\equiv D(p_i^2)$ and assuming we have $N$ data points, we minimized
\begin{equation}\label{tikdis1}
  \mathcal{J}_\lambda=\sum_{i=1}^{N}\left[\int_{\mu_0}^{+\infty} \d\mu\frac{\rho(\mu)}{p_i^2+\mu}-D_i\right]^2+\lambda \int_{\mu_0}^{+\infty} \d\mu~\rho^2(\mu)
\end{equation}
where we use lattice data in momentum space for the gluon propagator computed in a $80^4$ volume, with $\beta=6.0$ \cite{olisi12, letter}. The data was renormalized in a MOM scheme at $\mu=4$ GeV \cite{olisi12}. For $\lambda=0$, we would be searching that $\rho$ that reproduces the data as close as possible in norm. Though, we need $\lambda>0$ as a ``screening filter'' to overcome the ill-posed nature of the inversion. This amounts to Tikhonov regularization in a discrete setting. The Morozov principle amounts to fix the a priori free parameter $\lambda$ on that value $\overline\lambda$ whereby the quality of the inversion is equal to the error on the data, i.e.~
$||D^{reconstructed}-D^{data}||=\delta$ where $\delta$ is the total noise on the input data. We also introduced an IR regulator (threshold) $\mu_0$ into the game, the value thereof will be determined self-consistently by means of the optimal (Morozov) regulator $\overline\lambda$: we took the minimal value for $\overline\lambda(\mu_0)$ that can be reached by varying $\mu_0$. This sounds natural: the smaller $\lambda>0$ becomes the better we approach the original (ill-posed) problem.

Perturbing $\rho(\mu)$ linearly and demanding that the variation of $\mathcal{J}_\lambda$ vanishes, leads, after some manipulation, to the following equation that we need to solve for $\rho(\mu)$.
\begin{equation}\label{tikdis2}
  \sum_{i=1}^{N} \underbrace{\left[\int_{\mu_0}^{+\infty}\d\nu\frac{\rho(\nu)}{p_i^2+\nu}-D_i\right]}_{\equiv c_i}\frac{1}{p_i^2+\mu}+\lambda\rho(\mu)=0\,\, (\mu\geq\mu_0)
\end{equation}
Said otherwise, the (regularized) K\"all\'{e}n-Lehmann inverse is explicitly given by
\begin{equation}\label{tikdis3}
  \rho_{\lambda}(\mu)=-\frac{1}{\lambda}\sum_{i=1}^{N}\frac{c_i}{p_i^2+\mu}\theta(\mu-\mu_0)\,,
\end{equation}
with $\theta(\cdot)$ the Heaviside step function. Combination of eqns.~(\ref{tikdis2}) \& (\ref{tikdis3}) yields a linear system to be solved for the coefficients $c_i$:
\begin{equation}\label{tikdis5}
 \lambda^{-1} \mathcal{M}c+c=-D\,,
\end{equation}
with
\begin{equation}\label{tikdis6}
  \mathcal{M}_{ij}=\int_{\mu_0}^{+\infty}\d\nu\frac{1}{p_i^2+\nu}\frac{1}{p_j^2+\nu}=\frac{\ln\frac{p_j^2+\mu_0}{p_i^2+\mu_0}}{p_j^2-p_i^2}\,.
\end{equation}
The reconstructed propagator, which depends on $\lambda$, can be directly expressed as follows:
\begin{equation}\label{tikdis7}
  D^{reconstructed}(p^2)=\int_{\mu_0}^{+\infty}\d\mu\frac{\rho_{\lambda}(\mu)}{p^2+\mu}=-\frac{1}{\lambda}\sum_{i=1}^{N}\frac{c_i\ln\frac{p^2+\mu_0}{p_i^2+\mu_0}}{p^2-p_i^2}\,.
\end{equation}
The highest accessible lattice momenta reads $p_{max}=7.77~\textrm{GeV}$, a value that we will use here as an UV cut-off. The number of lattice data points was $124$ and the noise level comes as $\delta=0.658~\textrm{GeV}^{-2}$.

\begin{figure}[t] 
   \centering
\includegraphics[scale=0.27]{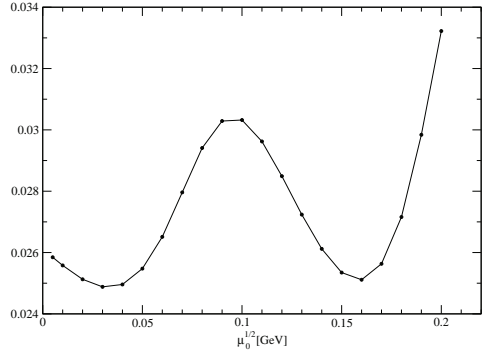}
  \caption{The Morozov parameter $\overline\lambda$ in terms of the threshold $\mu_0$.}
   \label{specT0-3}
\end{figure}

\begin{figure}[t] 
   \centering
   \subfigure[Spectral density.]{ \includegraphics[scale=0.3]{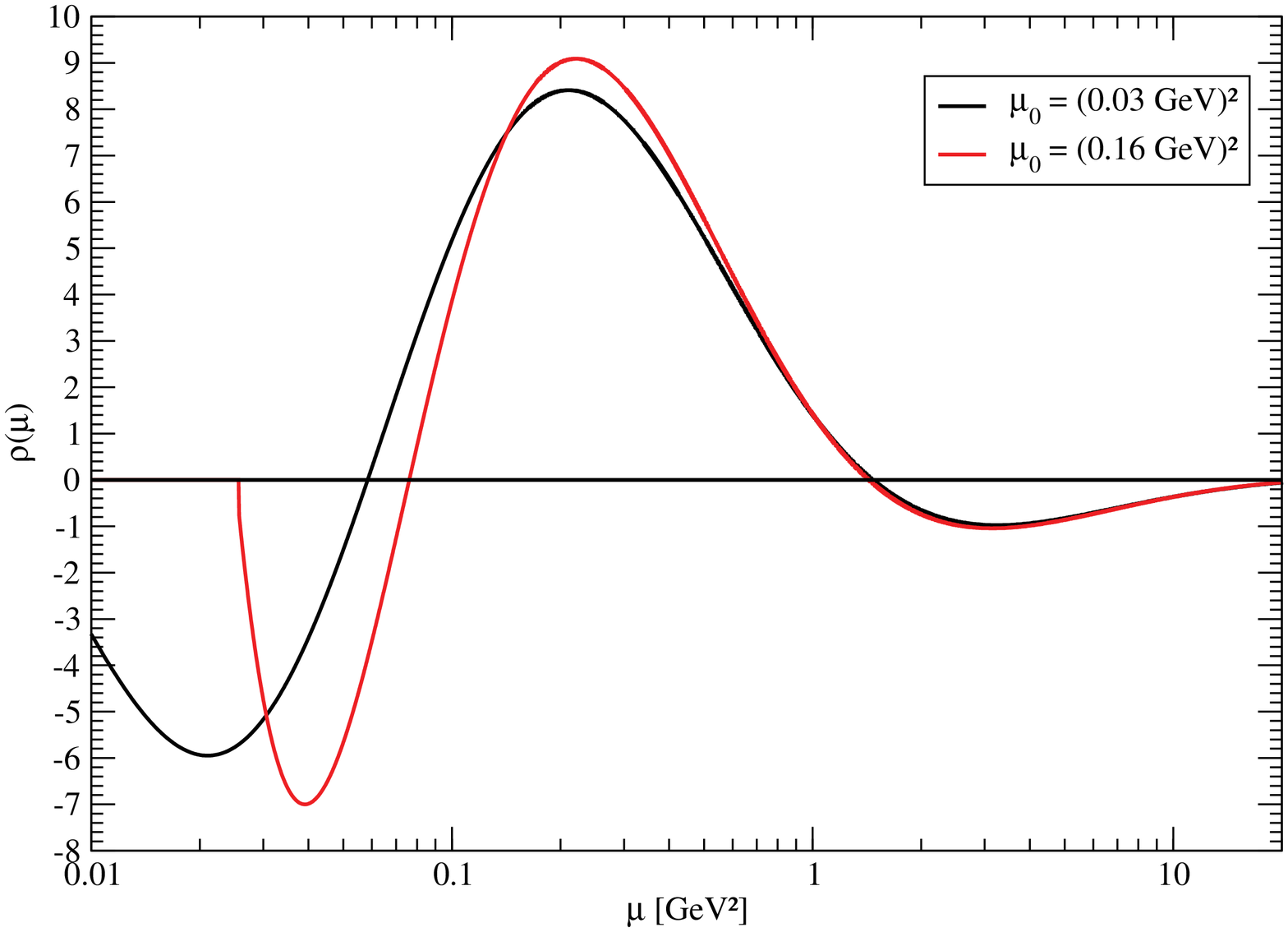} } \quad
   \subfigure[Reconstructed propagator.]{ \includegraphics[scale=0.3]{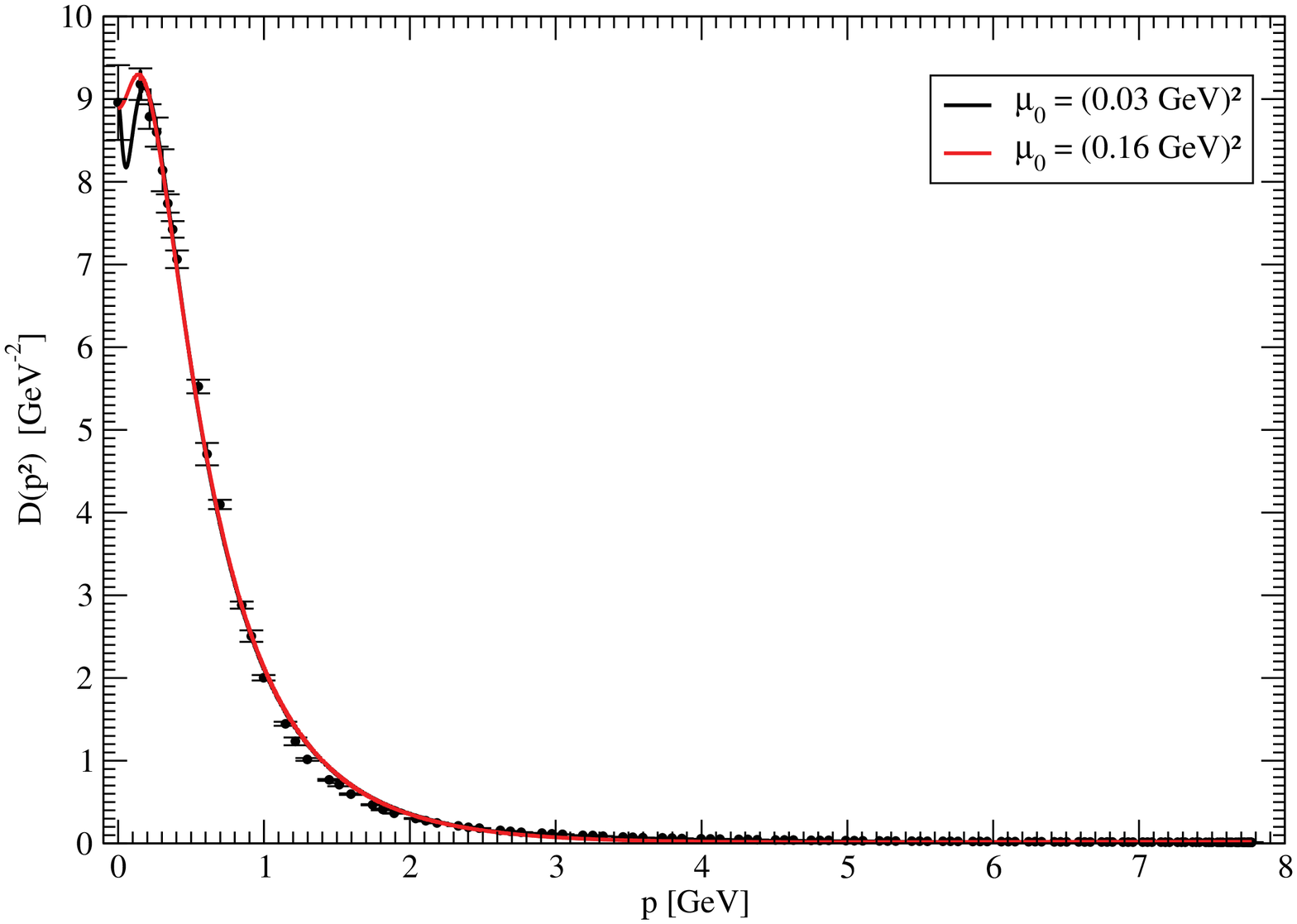} }
  \caption{Results for the gluon spectral function and the reconstructed propagator vs.~the input data. We refer to the main text and \cite{letter} for additional details.}
   \label{specT0-new}
\end{figure}
In Figure \ref{specT0-3} we notice the occurrence of 2 minima for $\overline\lambda(\mu_0)$, at $\mu_0\approx0.03~\textrm{GeV}^2$ and $\mu_0\approx0.16~\textrm{GeV}^2$, with the former one giving a slightly lower value of $\overline\lambda$. For both values, the reconstructed propagator and associated spectral density are shown in Figure \ref{specT0-new}. We clearly observe that the gluon spectral density is indeed a nonpositive quantity. One can also compare our estimate for the gluon spectral function, based on lattice data, with the numerical output of solving the complex momentum Dyson-Schwinger equations. With our current results, we do not see evidence of the reported sharp peak of \cite{fischer2012}, while the violation of positivity sets in well before $\omega\sim600~\textrm{MeV}$ \cite{fischer2012}.

\section{Gluons at finite temperature}

In this section, we consider lattice results for the gluon propagator at
finite temperature.  We study positivity violation through the computation
of the Schwinger and spectral functions, and also consider the interpretation
of the gluon propagator as a massive propagator.

The lattice setup for the simulations at finite temperature considered
here is described in Table \ref{tempsetup}. For further details see \cite{gluonmass}.

\begin{table}[t]
\begin{center}
\begin{tabular}{c@{\hspace{0.5cm}}c@{\hspace{0.3cm}}c@{\hspace{0.3cm}}r@{\hspace{0.4cm}}l@{\hspace{0.3cm}}l}
\hline
Temp.            & $\beta$ & $L_s$ &  \multicolumn{1}{c}{$L_t$} & \multicolumn{1}{c}{$a$} & \multicolumn{1}{c}{$1/a$} \\
 (MeV)           &              &            &            & \multicolumn{1}{c}{(fm)} & \multicolumn{1}{c}{(GeV)} \\
\hline
121 &   6.0000 & 64 & 16 & 	0.1016 &  	1.9426 \\
162 &   6.0000 & 64 & 12 & 	0.1016 & 	1.9426 \\
194 &   6.0000 & 64 & 10 & 	0.1016 & 	1.9426 \\
243 &   6.0000 & 64 &   8  & 	0.1016 & 	1.9426 \\
260 &   6.0347 & 68 &   8  & 	0.09502 & 	2.0767 \\
265 &   5.8876 & 52 &   6  & 	0.1243 & 	1.5881 \\
275 &   6.0684 & 72 &   8  & 	0.08974 & 	2.1989 \\
285 &   5.9266 & 56 &   6 & 	0.1154 & 	1.7103 \\
290 &   6.1009 & 76 &   8 & 	0.08502 & 	2.3211 \\
305 &   6.1326 & 80 &   8 & 	0.08077 & 	2.4432 \\
324 &   6.0000 & 64 &   6 & 	0.1016	 &      1.9426 \\
366 &   6.0684 & 72 &   6 & 	0.08974	 &      2.1989 \\
397 &   5.8876 & 52 &   4 & 	0.1243	 &      1.5881 \\
428 &   5.9266 & 56 &   4 & 	0.1154	 &      1.7103 \\
458 &   5.9640 & 60 &   4 & 	0.1077	 &      1.8324 \\
486 &   6.0000 & 64 & 	4 & 	0.1016	 &      1.9426 \\
\hline
\end{tabular}
\end{center}
\caption{Lattice setup used for the computation of the gluon propagator 
at finite temperature. Simulations used the Wilson gauge action; $\beta$ 
was adjusted to have a constant physical volume, $L_s \, a \simeq 6.5$ fm. 
For the generation of gauge configurations and Landau gauge fixing, we used 
Chroma \cite{chroma} and PFFT \cite{pfft} libraries.}
\label{tempsetup}
\end{table}

\subsection{Positivity violation and spectral densities}

\begin{figure}[b]
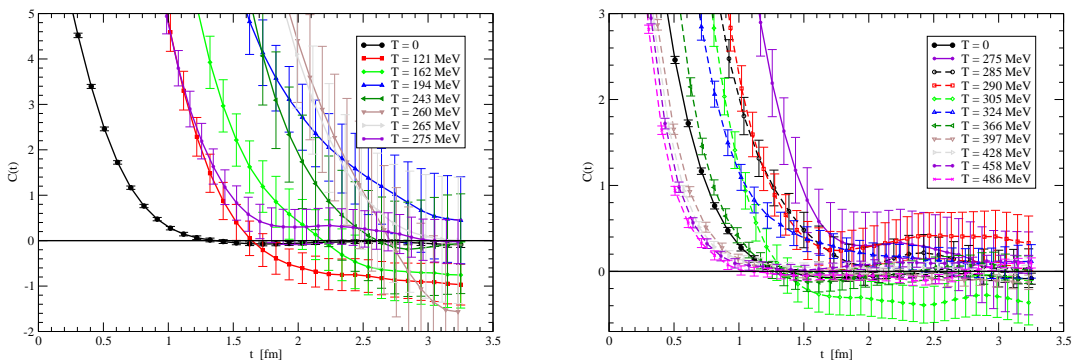
 
\vspace*{0.3cm}
   \centering
   \subfigure{ \includegraphics[scale=0.28]{plots/spectral/Tfinita/positividade_long_upTc} } \qquad
   \subfigure{ \includegraphics[scale=0.28]{plots/spectral/Tfinita/positividade_long_aboveTc} }
  \caption{Temporal correlator for the longitudinal component.}
   \label{poslong}
\end{figure}

\begin{figure}[t]
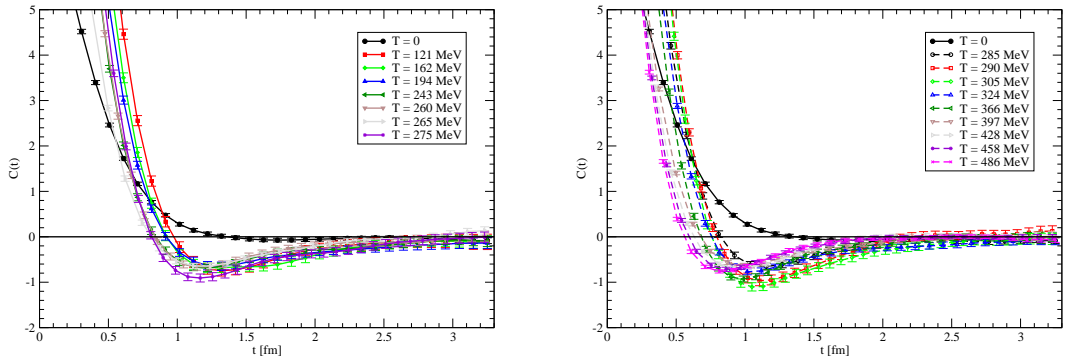
 
\vspace*{0.6cm}
   \centering
   \subfigure{ \includegraphics[scale=0.28]{plots/spectral/Tfinita/positividade_trans_upTc} } \qquad
   \subfigure{ \includegraphics[scale=0.28]{plots/spectral/Tfinita/positividade_trans_aboveTc} }
  \caption{Temporal correlator for the transverse component.}
   \label{postrans}
\end{figure}

\begin{figure}[h]
\vspace*{0.5cm}
\begin{center}
\includegraphics[width=0.5\textwidth]{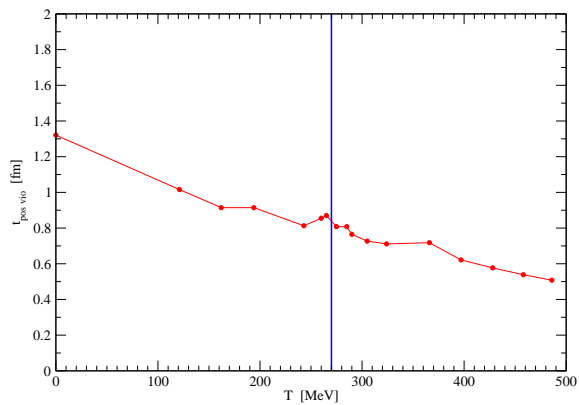}
\caption{Positivity violation scale for the transverse propagator.}
\label{viotrans}
\end{center}
\end{figure}

In this subsection, the temporal correlator defined in eq.~(\ref{schwinger}) is
computed for the longitudinal and transverse components of the gluon propagator
for the temperatures described in Table \ref{tempsetup}.
The Schwinger function for the longitudinal form factor, see Figure \ref{poslong}, and
for the transverse form factor, see Figure \ref{postrans}, show that positivity is
violated for both transverse and longitudinal components at all temperatures.
From the results for the transverse propagator, we see that the time scale for
positivity violation decreases with the temperature -- see Figure \ref{viotrans}.
This suggests that, for sufficiently high temperatures, transverse 
gluons can behave as quasi-particles. In what concerns the behaviour of the 
time scale associated with the violation of positivity for
the longitudinal propagator with the temperature, the results are not so clear 
as for the transverse Schwinger function.

\begin{figure}[t]
\vspace*{0.5cm}
\begin{center}
\includegraphics[width=0.6\textwidth]{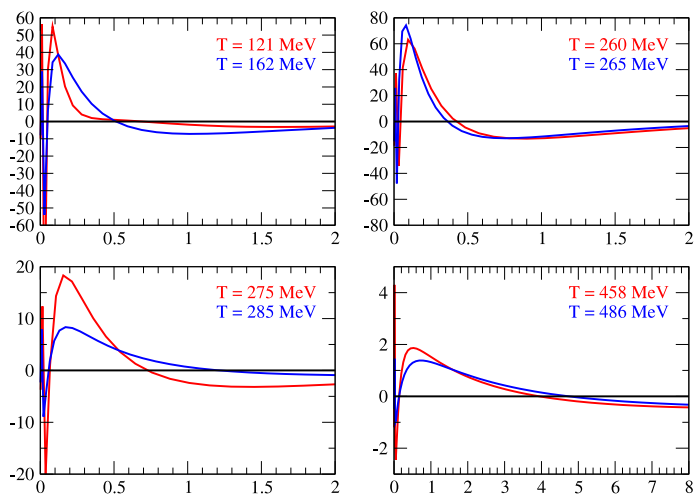}
\caption{Longitudinal propagator spectral densities.}
\label{specT}
\end{center}
\end{figure}

The spectral densities\footnote{For another recent work concerning gluon spectral functions at finite temperature see \cite{haas}.} associated with the longitudinal gluon form factor 
can be seen in Figure \ref{specT}.  The spectral density $\rho ( \mu )$ is negative for large $\mu$ and
the energy scale at which the $\rho ( \mu )$ becomes negative seems to increase with temperature.
This, again, suggests that for sufficiently high temperatures, longitudinal gluons may be considered as
massive quasi-particles.

\subsection{Gluon mass}

\begin{figure}[h]
\vspace*{0.3cm}
\begin{center}
\includegraphics[width=0.5\textwidth]{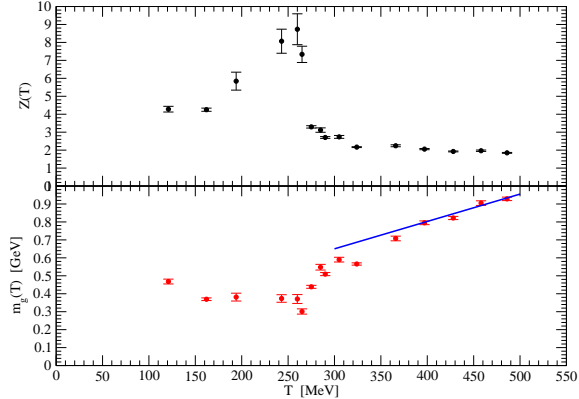}
\caption{$Z(T)$ and $m_g(T)$ from fitting the longitudinal
gluon propagator to a Yukawa form.
The curve in the lower plot is
the fit of $m_g$ to the functional form predicted by perturbation theory.}
\label{yukawafigure}
\end{center}
\end{figure}

One can associate with the gluon propagator a mass scale. Here we consider different definitions for an
electric (longitudinal) and magnetic (transverse) gluon mass scale as a function of the temperature -- see
\cite{gluonmass} for details.

In a quasi-particle picture, the gluon is considered a massive boson 
with a Yukawa-type propagator
\begin{equation}
 D(p) = \frac{Z }{ p^2 + m^2} \ ,
\label{yukawa}
\end{equation}
where $m$ is the gluon mass and $Z^{\frac{1}{2}}$ the overlap between 
the gluon state and the quasi-particle massive state.

A value for the gluon mass can be obtained fitting the infrared lattice data to  eq. (\ref{yukawa}).
In Table \ref{yukawatable} and in Figure \ref{yukawafigure} we report the functions
$Z(T)$ and $m_g (T)$ associated with the longitudinal component of the propagator.
In what concerns the transverse form factor, it turns out that it is not described by a Yukawa-like function
in the infrared region and one concludes that the magnetic propagator
does not behave as a quasi-particle massive boson for
$T \lesssim 500$ MeV.

\begin{table}[!t]
\begin{center}
\begin{tabular}{c@{\hspace{0.5cm}}l@{\hspace{0.5cm}}l@{\hspace{0.5cm}}l@{\hspace{0.4cm}}l}
\hline
 Temp.           & $p_{max}$ & $Z(T)$            &  $m_g(T)$ & \multicolumn{1}{c}{$\chi^2/d.o.f.$} \\
 (MeV)          &   (GeV)                 &                   &   (GeV)       &  \\
 \hline
 121              & 0.467          & 4.28(16)    & 0.468(13)                                   &   1.91 \\
 162              & 0.570          &  4.252(89) & 0.3695(73)                                 &   1.66 \\
 194              & 0.330          &  5.84(50)   & 0.381(22)                                   &   0.72 \\
 243              &  0.330          &  8.07(67)  & 0.374(21)                                   &   0.27 \\
 260              &  0.271          &  8.73(86)  & 0.371(25)                                   &   0.03 \\
 265              &  0.332          &  7.34(45)  & 0.301(14)                                   &   1.03  \\
 275              &  0.635          &  3.294(65) &  0.4386(83)                               &   1.64  \\
 285              &  0.542          &  3.12(12)   & 0.548(16)                                  &   0.76  \\
 290              &  0.690          &  2.705(50) & 0.5095(85)                                &   1.40  \\
 305              &  0.606          &  2.737(80) & 0.5900(32)                                &   1.30 \\
 324              &  0.870          &  2.168(24) & 0.5656(63)                                &   1.36 \\
 366              &  0.716          &  2.242(55) & 0.708(13)                                  &   1.80 \\
 397              &  0.896          &  2.058(34) & 0.795(11)                                   &  1.03 \\
 428              &  1.112          &  1.927(24)  & 0.8220(89)                                &  1.30 \\
 458              & 0.935          &  1.967(37)  & 0.905(13)                                   & 1.45  \\
 486              & 1.214          &  1.847(24)  & 0.9285(97)                                 & 1.55  \\
\hline
\end{tabular}
\end{center}
\caption{Results of fitting the longitudinal propagator $D_{L}(p^{2})$ to a Yukawa form from $p = 0$ up to $p_{max}$.}
\label{yukawatable}
\end{table}

\begin{figure}[h] 
\vspace*{0.6cm}
   \centering
   \subfigure[Our results.]{ \includegraphics[scale=0.28]{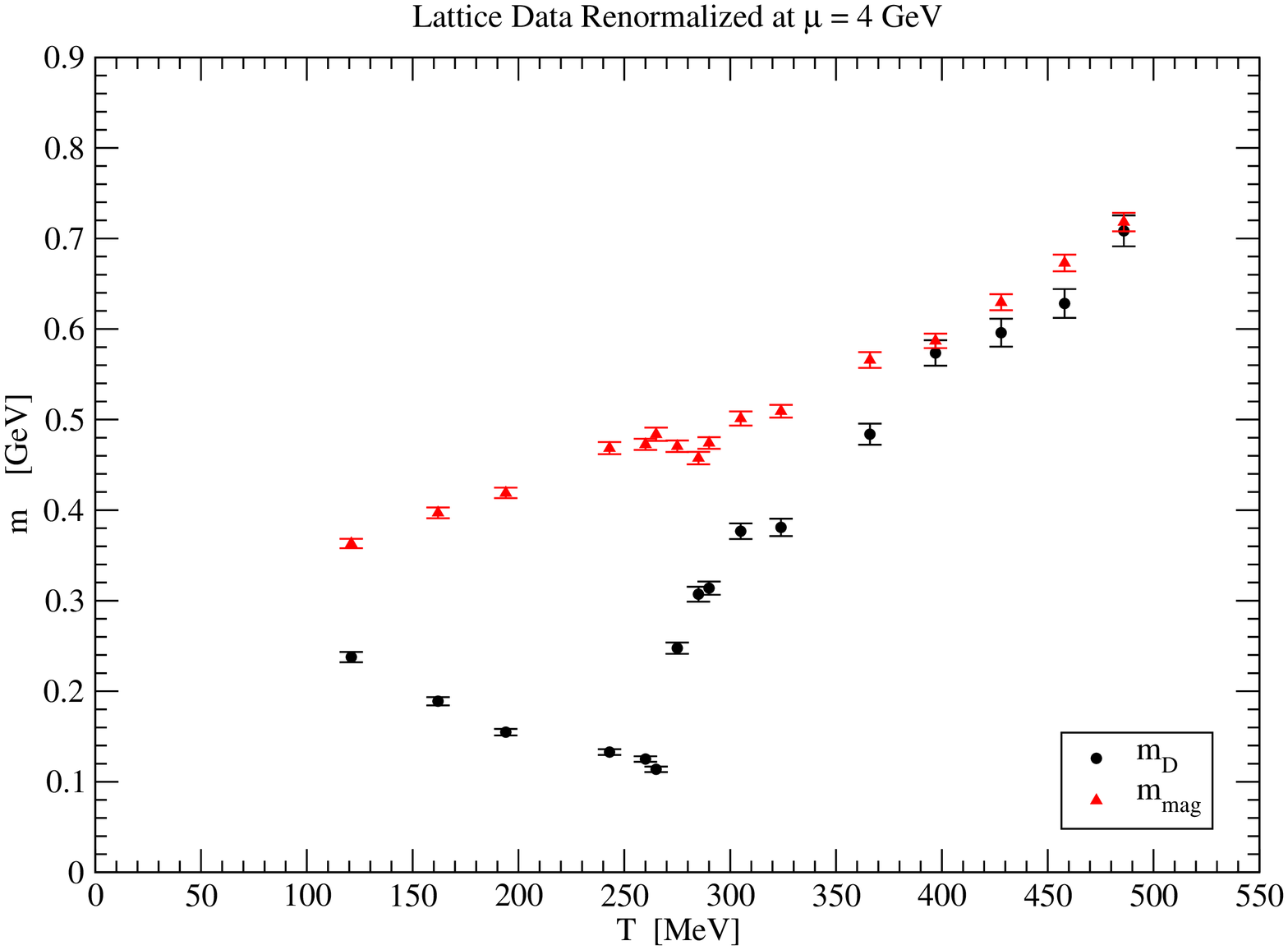} } \qquad
   \subfigure[Comparison with \cite{maas2012}.]{ \includegraphics[scale=0.28]{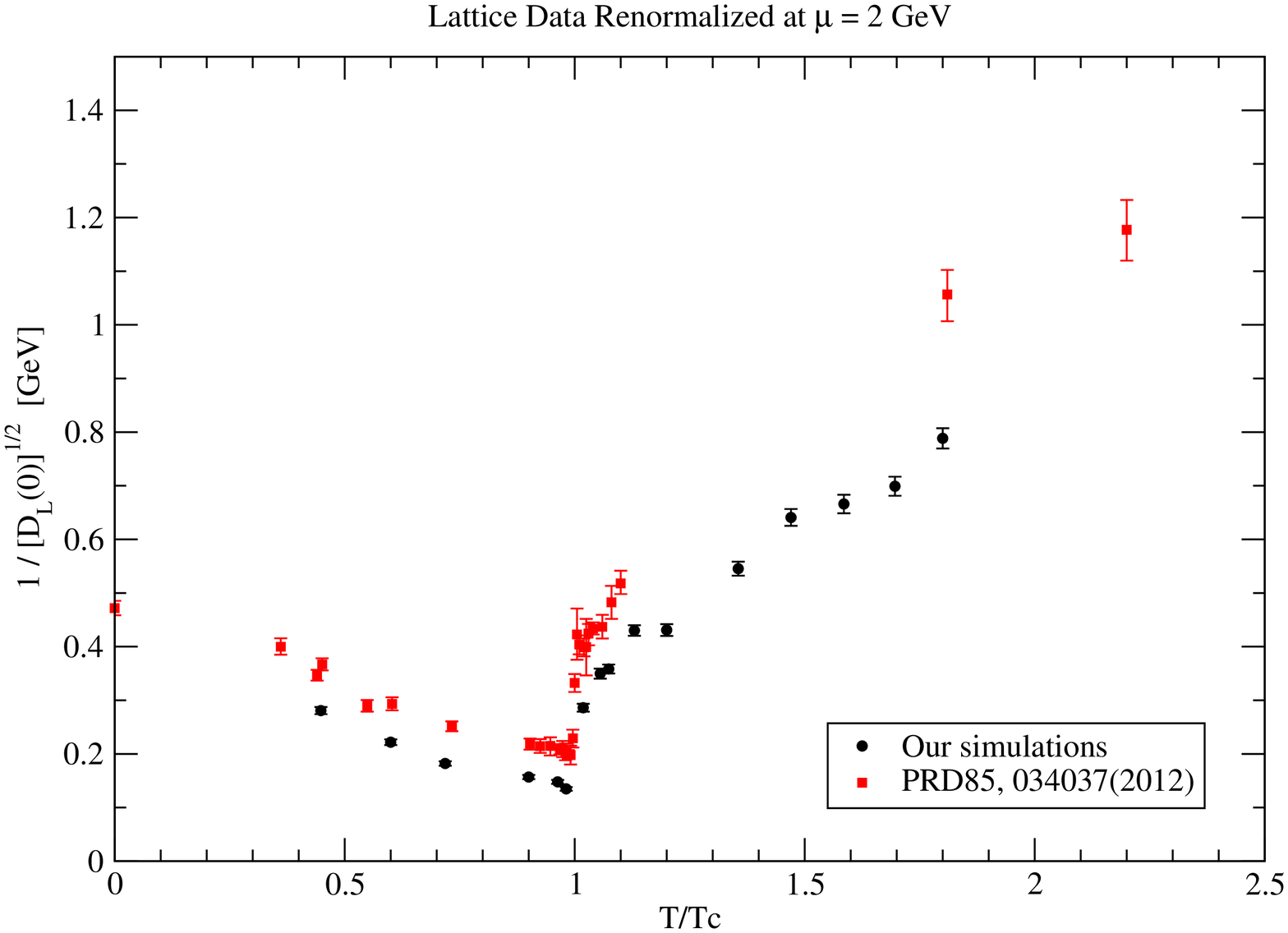} }
\caption{Electric and magnetic mass defined from zero momentum propagators. }
   \label{massD0}
\end{figure}

Another nonperturbative mass scale associated with gluon propagator 
that can be defined is given by
\begin{equation}
 m = 1 / \sqrt{ D(p^2=0; T) } \ .
\label{zeromass}
\end{equation}
Such a mass scale is reported in Figure \ref{massD0} for both electric and 
magnetic components. Moreover, in the right-hand plot we compare our data for 
the electric component with the results given in \cite{maas2012} using 
the same renormalization condition.

\section*{Acknowledgments}
Work supported by FCT via projects
CERN/FP/123612/2011, CERN/FP/123620/2011, and PTDC/FIS/100968/2008,
developed under the initiative QREN financed by the UE/FEDER through
the Programme COMPETE - Programa Operacional Factores de Competitividade.
P.~J.~Silva supported by FCT grant SFRH/BPD/40998/2007.
D.~Dudal acknowledges financial support from the Research-Foundation
Flanders (FWO Vlaanderen) via the Odysseus grant of F.~Verstraete.


\newpage

\begin{thebibliography}{99}
\bibitem{cucc07}
A.~Cucchieri and T.~Mendes, \pos{PoS(LATTICE 2007)297}.

\bibitem{bma09}
I.~L.~Bogolubsky, E.~M.~Ilgenfritz, M.~Muller-Preussker and A.~Sternbeck, Phys.\ Lett.\ B {\bf 676} (2009) 69.

\bibitem{olisi12}
O.~Oliveira and P.~J.~Silva,  Phys.\ Rev.\ D {\bf 86} (2012) 114513.

\bibitem{SilOli04}
P.~J.~Silva, O.~Oliveira, Nucl. Phys. B \textbf{690} (2004) 177-198.

\bibitem{Stern13}
A.~Sternbeck, M.~M\"{u}ller-Preussker, Phys. Lett. B \textbf{726} (2013) 396-403.

\bibitem{lit} V.~G.~Bornyakov, E.-M.~Ilgenfritz, C.~Litwinski, V.~K.~Mitrjushkin, M.~M\"{u}ller-Preussker, {\tt arXiv:1302.5943 [hep-lat]}.

\bibitem{cucc05}
A.~Cucchieri, T.~Mendes and A.~R.~Taurines, Phys.\ Rev.\ D {\bf 71} (2005) 051902.

\bibitem{aubin04}
C.~Aubin and M.~C.~Ogilvie, Phys.\ Rev.\ D {\bf 70} (2004) 074514.

\bibitem{sioli06}
P.~J.~Silva and O.~Oliveira, \pos{PoS(LAT2006)075}.

\bibitem{bowman07} P.~O.~Bowman, U.~M.~Heller, D.~B.~Leinweber, M.~B.~Parappilly, A.~Sternbeck, L.~von~Smekal, A.~G.~Williams and J.-b.~Zhang, Phys.\ Rev.\ D {\bf 76} (2007) 094505.

\bibitem{cornwall} J.~M.~Cornwall, {\tt arXiv:1310.7897 [hep-ph]}.

\bibitem{mem}
M.~Asakawa, T.~Hatsuda and Y.~Nakahara, Prog.\ Part.\ Nucl.\ Phys.\  {\bf 46} (2001) 459.

\bibitem{latt2012}
O.~Oliveira, D.~Dudal and P.~J.~Silva, \pos{PoS(Lattice 2012)214}.

\bibitem{qchsx}
D.~Dudal, P.~J.~Silva and O.~Oliveira, \pos{PoS(Confinement X)033}.

\bibitem{latt2013} P.~J.~Silva, D.~Dudal, O.~Oliveira, \pos{PoS(LATTICE 2013)366}, {\tt arXiv:1311.3643 [hep-lat]}.

\bibitem{letter}
D.~Dudal, O.~Oliveira and P.~J.~Silva, Phys.\ Rev.\ D, to appear, {\tt arXiv:1310.4069 [hep-lat]}.

\bibitem{gluonmass}
P.~J.~Silva, O.~Oliveira, P.~Bicudo and N.~Cardoso, {\tt arXiv:1310.5629 [hep-lat]}.

\bibitem{chroma} R.~G.~Edwards and B.~Jo\'o,
Nucl.Phys.Proc.Suppl. {\bf 140} (2005) 832 [{\tt arXiv:hep-lat/0409003 [hep-lat]}].

\bibitem{pfft} M.~Pippig, SIAM J. Sci. Comput. {\bf 35}, C213 (2013).

\bibitem{fischer2012}
S.~Strauss, C.~S.~Fischer and C.~Kellermann, Phys.\ Rev.\ Lett.\  {\bf 109} (2012) 252001.

\bibitem{haas}
M.~Haas, L.~Fister, J.~M.~Pawlowski, {\tt arXiv:1308.4960 [hep-ph]}.

\bibitem{maas2012} A.~Maas, J.~M.~Pawlowski, L.~von~Smekal, and D.~Spielmann,
Phys.\ Rev.\ D {\bf 85} (2012) 034037.



\end{thebibliography}
\end{document}